\documentclass[12pt,preprint]{aastex}
\usepackage{natbib}

\begin{document}

\title{Principal Component Analysis of SDSS Stellar Spectra}
\author{Rosalie C. McGurk, Amy E. Kimball, \v{Z}eljko Ivezi\'{c}}

\affil{Department of Astronomy, University of Washington, 
        Box 351580, Seattle, WA 98195}

\begin{abstract}
We apply Principal Component Analysis (PCA) to $\sim$100,000 stellar 
spectra obtained by the Sloan Digital Sky Survey (SDSS). In order to 
avoid strong non-linear variation of spectra with effective temperature, 
the sample is binned into 0.02 mag wide intervals of the $g-r$ color 
($-0.20<g-r<0.90$, roughly corresponding to MK spectral types A3 to K3),
and PCA is applied independently for each bin. In each color bin, the 
first four eigenspectra are sufficient to describe the observed spectra 
within the measurement noise. We discuss correlations of 
eigencoefficients with metallicity and gravity estimated by the 
Sloan Extension for Galactic Understanding and Exploration 
(SEGUE) Stellar Parameters Pipeline. 
The resulting high signal-to-noise mean spectra and the other 
three eigenspectra are made publicly available. 
These data can be used to generate high quality spectra for an 
arbitrary combination of effective temperature, metallicity, and 
gravity within the parameter space probed by the SDSS.
The SDSS stellar spectroscopic database and the PCA results presented 
here offer a convenient method to classify new spectra, to search 
for unusual spectra, to train various spectral classification methods, 
and to synthesize accurate colors in arbitrary optical bandpasses. 
\end{abstract}


\section{Introduction}

A large number of homogeneously-obtained stellar spectra have recently 
become available. For example, the Sloan Digital Sky Survey (SDSS) 
\citep{York.etal.2000AJ} has made publicly 
available\footnote{See http://www.sdss.org/dr7} over 460,000 stellar 
spectra as a part of its Data Release 7 \citep{Abazajian.2008arXiv}, 
and Radial Velocity 
Experiments\footnote{See http://www.rave-survey.aip.de/rave} (RAVE) may 
provide up to a million spectra over the next few years. This rapid 
progress in the availability of stellar spectra re-opens the old question 
of optimal stellar parameter extraction. For example, the SDSS estimates
effective temperature, gravity, and metallicity using a variety of 
standard methods implemented in an automated pipeline (SEGUE\footnote{Sloan 
Extension for Galactic Understanding and Exploration} Stellar 
Parameters Pipeline, hereafter SSPP; \citealt{2006MmSAI..77.1171B}). 
A detailed discussion of these methods and their performance can be 
found in \citet{2006ApJ...636..804A, 2007arXiv0710.5780A} and 
\citet{2007arXiv0710.5645L,2007arXiv0710.5778L}. The results of 
different methods implemented in the SSPP are {\it averaged} to obtain the 
final adopted values in the SDSS Spectral Parameter Pipeline table 
({\it sppParams}). Although a 
detailed analysis by \citet{2007arXiv0710.5645L, 2007arXiv0710.5778L} 
demonstrates that systematic metallicity differences between the methods 
used in averaging do not exceed $\sim$0.1 dex (with random errors
in the range 0.1--0.3 dex), it is fair to ask whether a single method 
could be used to obtain the same level of systematic and random errors, 
instead of combining different methods with varying error properties.

Principal Component Analysis (PCA) has been demonstrated as a viable tool 
in solving this classification problem 
(\citealt{1995AJ....110.1071C, 1998IAUS..179..376C, 1998MNRAS.298..361B}; 
and references therein). \citet{2004AJ....128.2603Y} have developed a 
PCA-based analysis 
code specialized to SDSS spectra. Here we use the same code to 
investigate whether the PCA eigencoefficients are correlated with 
the metallicity and gravity obtained by the SSPP. Byproducts of this
analysis are high signal-to-noise eigenspectra that can be used to 
generate spectra for any combination
of basic stellar parameters (effective temperature, metallicity, 
and gravity) within the parameter space probed by SDSS. Hence,
given an arbitrary spectrum, one can attempt a low-dimensional
fit using our library of eigenspectra. Among numerous drivers for such 
a library, we single out a photometric calibration scheme for the 
Large Synoptic Survey Telescope (LSST)\footnote{See http://www.lsst.org/}.  
LSST plans to use an auxiliary spectroscopic telescope to obtain 
spectra of standard stars at the same time as the main imaging 
survey is performed \citep[see][]{2008arXiv0805.2366I}. 
The atmospheric transmission properties, required to photometrically 
calibrate the imaging survey, will be obtained by simultaneously 
fitting the stellar spectrum and a sophisticated atmospheric model 
with six free parameters for each observation. The ability to 
describe the expected stellar spectra in a low-dimensional continuous 
space by using a small number of eigencomponents, with eigencoefficients 
that are not defined on a fixed grid, might increase the fidelity 
of the fitted model. 

In Section 2 we describe our sample selection and the application
of PCA to SDSS stellar spectra. We discuss our results in Section 3,
and end with a summary in Section 4.

\section{Principal Component Decomposition of SDSS Stellar Spectra }

\subsection{ The properties of SDSS spectra  }
In addition to massive amounts of optical photometry of unprecedented 
quality, the SDSS has 
also produced a large spectroscopic database. A compendium of technical details 
about SDSS can be found on the SDSS web site\footnote{See http://www.sdss.org/}, 
which also provides an interface for public data access. Targets for the 
spectroscopic survey are chosen from the SDSS imaging data based on their 
colors and morphological properties \citep{Strauss.2002AJ, Eisenstein.2001AJ, Richards.2002AJ}. In the spectroscopic survey, stars are 
targeted either as calibrators or for scientific reasons in specific 
parts of the four-dimensional SDSS color space 
\citep{Yanny09}.

A pair of multi-object fiber-fed spectrographs mounted onto the 
SDSS 2.5m telescope \citep{2006AJ....131.2332G} are used to take 640 
simultaneous spectra within a radius of 1.49 degrees, each with 
wavelength coverage 3800--9200~\AA~and spectral resolution of 
$\sim$$2000$, and with a signal-to-noise ratio of $>$4 per pixel at 
$g$=20.2.  Spectro-photometric calibration of these spectra is 
exquisite;  for example, 
the imaging magnitudes and the stellar magnitudes synthesized from SDSS spectra 
agree with an rms of only $\sim$0.05 mag \citep[see][]{2004ApJ...615L.141S}.

\subsection{ Sample selection } %% 2.2
\label{sample}
We begin by selecting bright stars in SDSS Data Release 6 that have colors
consistent with the main stellar locus 
\citep{1998ApJS..119..121L, 1999AJ....117.2528F, 2000AJ....120.2615F}, 
or are found in the regions populated by RR Lyrae stars \citep{2005AJ....129.1096I} 
and blue horizontal branch stars \citep{2004AJ....127..899S}. Stars that 
are probable white dwarf --- red dwarf pairs \citep{2004ApJ...615L.141S} or 
single hot white dwarfs \citep{2006ApJS..167...40E} are not selected. We only use 
stars from the sky regions with modest interstellar dust extinction, determined 
using the interstellar dust maps of \citet{1998ApJ...500..525S}. 

The specific criteria applied to 130,620 entries from the SDSS DR6 version of
{\it sppParams} 
table\footnote{See http://www.sdss.org/dr6/products/spectra/spectroparameters.html} 
that have $\log(g)>0$ are (the number in brackets indicates the number of
stars remaining after each selection step):

\begin{enumerate}
\item the interstellar extinction in the $r$ band below 0.3; [106,816]
\item $14 < g < 19.5$; [104,844]
\item $-0.2 < g-r < 0.9$; [103,588]
\item $ 0.7 < u-g < 2.4$; [101,630]
\item {\bf \{}$ -0.2 < g-r - 0.5(u-g-0.5) < 0.4 ${\bf \}} OR {\bf \{}$u-g < 1.4$ AND $g-r < 0.25${\bf \}}; [100,759]
\item $-0.2 < 0.35(g-r) - (r-i) < 0.20$; [98,063]
\end{enumerate}

\begin{figure}
\plotone{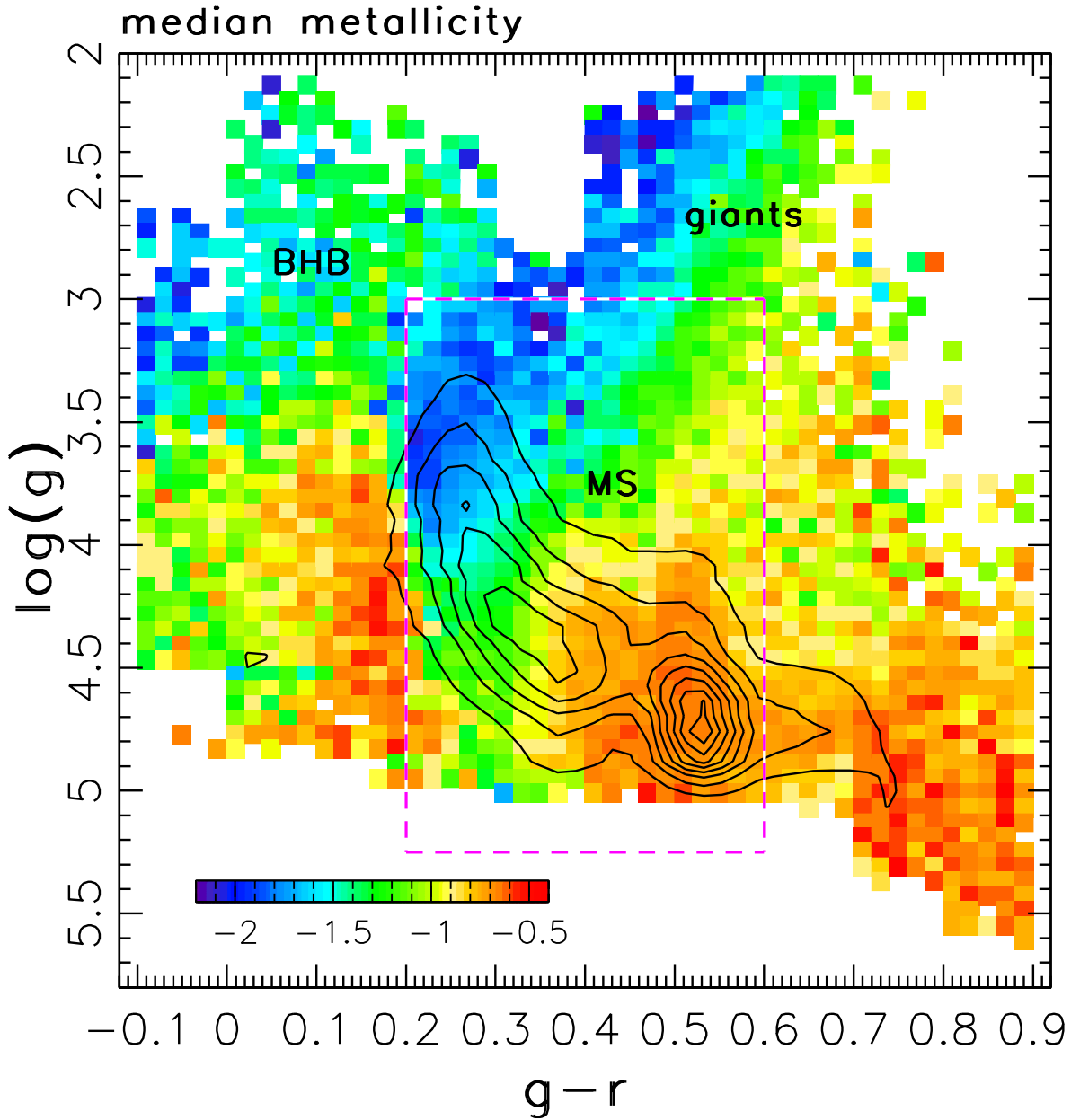} 
\caption{
The linearly-spaced contours show the distribution of $\sim$100,000 stars with
$g<19.5$ from the SDSS DR6 spectroscopic sample in the $\log(g)$ vs.~$g-r$ plane. 
The multi-modal distribution is a result of the SDSS target selection algorithm. 
The color scheme shows the median metallicity in all 0.02 mag by 0.06
dex large pixels that contain at least 10 stars (according to the legend 
shown in the bottom left corner). The fraction of stars with
$\log(g)<3$ (giants) is 4\%, and they are mostly found in two color regions:
$-0.1 < g-r < 0.2$ (BHB stars) and $0.4 < g-r < 0.65$ (red giants). They are
dominated by low-metallicity stars ($[Fe/H]<-1$). The dashed lines roughly 
outline the main-sequence (MS) region, see \citet{2008arXiv0804.3850I}.}
\label{Fig:logg}
\end{figure}

For each star, the data analyzed in this work include the $ugriz$ photometry,
SDSS spectrum, and SSPP estimates of effective temperature ($T_{eff}$), 
metallicity ($[Fe/H]$), and gravity ($\log(g)$). The selected stars span the 
range of effective temperature from $\sim$4500 K to $\sim$9000 K (see below), 
and 99.4\% have metallicity in the range $-3<$[Fe/H]$<0$ with a median of $-1.0$. 
While the sample is dominated by main sequence stars (the median $\log(g)$ is 4.1 
$\pm$ 0.44 dex), a small fraction of stars ($\sim$3\%) have lower
gravity estimates consistent with giants (see Figure~\ref{Fig:logg}).

\subsection{ The $g-r$ Color Binning } %% 2.3

\begin{figure}
\vskip -0.5in
\epsscale{0.9}
\plotone{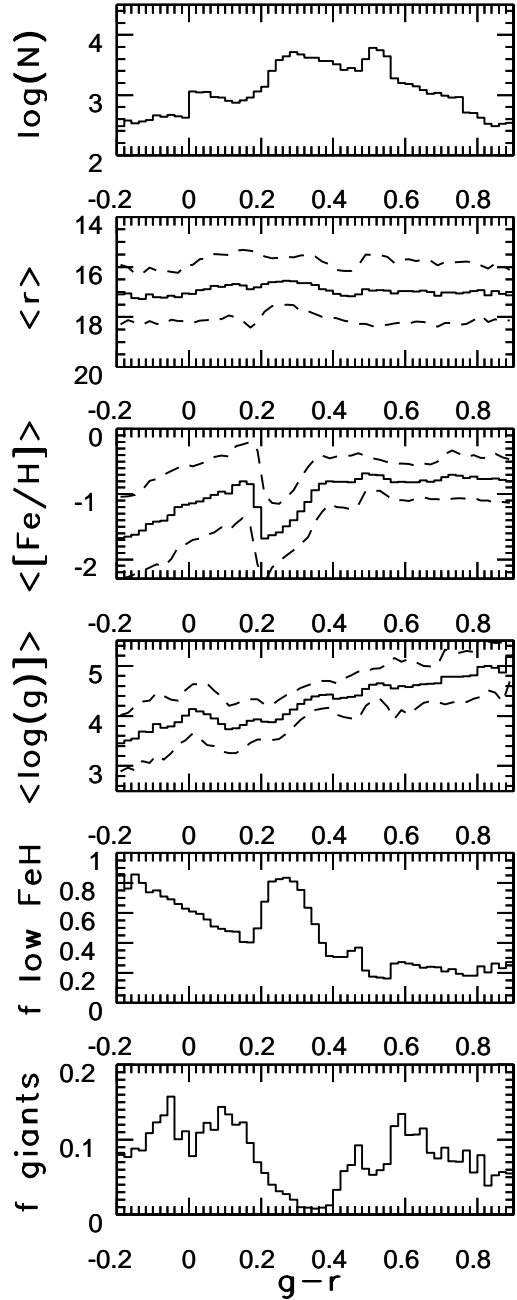}
\epsscale{1.0}
\vskip -0.4in
\caption{Top to bottom: The solid lines show the variation of the number of 
stars in a given $g-r$ bin, the median apparent $r$ band magnitude, 
metallicity, gravity, and the fraction of low-metallicity stars ($[Fe/H]<-1$) 
and giants ($\log(g)<3$). Dashed lines show 
the 1$\sigma$ envelope around the medians. Stars with $g-r>0.2$ are dominated
by main sequence stars, and the bluer stars are dominated by blue horizontal
branch stars, RR Lyrae stars, and blue stragglers. The approximate $g-r$ 
colors for several MK spectral types (luminosity class V) are taken from 
\citet{Covey.2007AJ}.}
\label{Fig:grStats}
\end{figure}

Stellar spectra are to the zero-th order similar to the Planck 
(black body) function controlled by the effective temperature; however, their 
variation of spectral line width and strength depend not only on 
effective temperature but also on metallicity and gravity.  At a 
chosen effective temperature, the metallicity affects the strength of 
the spectral line and gravity can broaden or narrow the spectral line.
In order to study these variations due to metallicity and gravity at a 
given effective temperature, we group the stars with SDSS spectra 
into 55 color bins with a width of 0.02 mag, in the range $-0.2 < g-r < 0.9$. 
As shown by \citet{2008arXiv0804.3850I}, this color is strongly 
correlated with the effective temperature determined by the SSPP: 0.02 mag 
wide bin in the $g-r$ color roughly corresponds to one MK spectral 
subtype. The best-fit expression derived by \citet{2008arXiv0804.3850I}, 
\begin{equation}
\label{logT}
   \log(T_{\rm eff} / {\rm K}) = 
          3.882 - 0.316(g-r) + 0.0488(g-r)^2 + 0.0283(g-r)^3,
\end{equation}
achieves systematic errors below 0.004 dex and overall rms of 0.008 dex,
within the $-0.3 < g-r < 1.3$ color range. The temperature range 
corresponding to the $g-r$ limits adopted here ($-0.2 < g-r < 0.9$)
is 4,550 K -- 8,850 K. The number of stars per $g-r$ bin ranges from 
430 to 9104, with  a median of 1571. The variation of the number
of stars in a bin, the median apparent magnitude, metallicity, gravity, 
and the fractions of low-metallicity stars ($[Fe/H]<-1$) and giants
($\log(g)<3$) are shown in Figure~\ref{Fig:grStats}. 

\begin{figure}
\epsscale{0.8}
\plotone{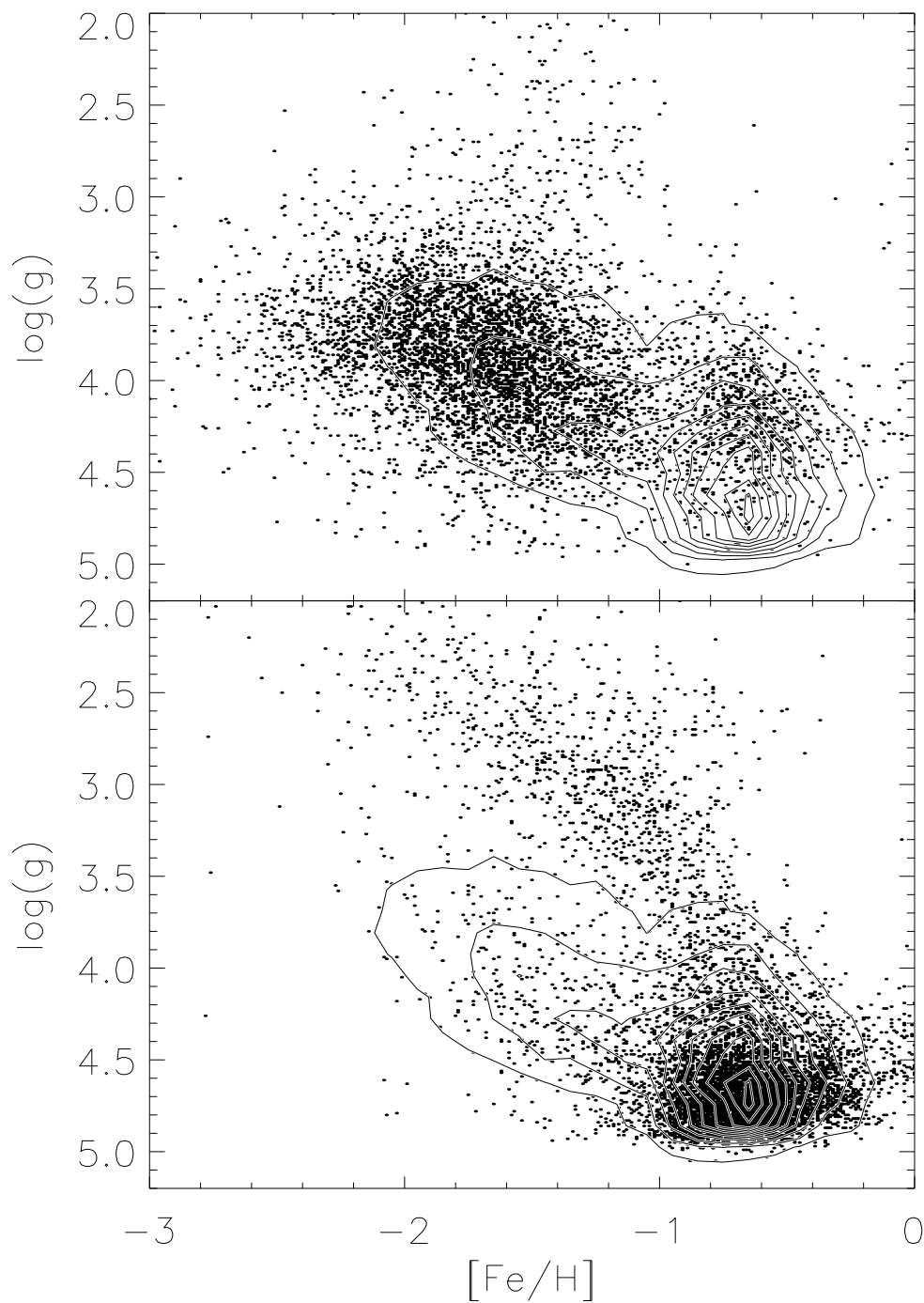}
\vskip -0.35in
\caption{ 
The bivariate metallicity-gravity distribution for two color bins. 
In both panels, the linearly-spaced contours show the $\log(g)$ vs. 
$[Fe/H]$ distribution of the $\sim$100,000 stars in the analyzed 
sample. The symbols in the top panel show stars from bin \#23, and 
the bottom panel shows stars from bin \#37. Bin \#37 ($\log(g)<3.5$ 
and $[Fe/H]<-1$) contains many low-metallicity giant stars.}
\label{Fig:FeHLogg}
\end{figure}

Most of the stars in our sample are main-sequence stars with 
$\log(g)>3$ and $[Fe/H] > -1$. The impact of metallicity and gravity 
on stellar spectra, as well as the fraction of stars in the sample
that are not main-sequence disk stars, varies with effective temperature,
i.e., with the $g-r$ color. For the purposes of presentation,
we single out two bins in $g-r$. For the 6156 stars in bin \#23 
($0.24 < g-r < 0.26$) we expect a strong correlation between the $u-g$ 
color and metallicity, as discussed in detail by \cite{2008arXiv0804.3850I}. 
We use this bin to compare the random errors for metallicity estimates
obtained from the SSPP and obtained here using PCA. For bin \#37, with 
$0.52 < g-r < 0.54$ and 8253 stars, the fraction of giants is near 
its maximum ($\sim$10\%).  This bin enables a study of correlations 
between log(g) and PCA eigencoefficients. 
The color range of bin \#37 was deliberately targeted for SDSS 
spectroscopy because giant stars are good probes of distant halo
structure. The distribution of stars from these two bins in the
$\log(g)$ vs. $[Fe/H]$ metallicity diagram is compared to the full
sample in Figure~\ref{Fig:FeHLogg}.

\subsection{Principal Components Decomposition} %% 2.4

For a thorough discussion of PCA and several of its various 
applications, we refer the reader to \citet{1995AJ....110.1071C} 
and \citet{2004AJ....128.2603Y}. Briefly, PCA calculates 
eigenspectra, or characteristic averaged spectral shapes, from the input 
batch of spectra, and measures eigencoefficients that represent how 
strongly each eigenspectrum is present in a data spectrum. The PCA package 
we used was specifically developed for use with SDSS spectra 
\citep{2004AJ....128.2603Y}.  Before PCA decomposition occurs, the package 
shifts the spectra to their rest-frames (zero radial velocity) and rebins them 
to a common wavelength range.  The spectra are then repaired in gappy regions 
(i.e. bad pixels or missing data) using the iterative KL-correction formalism 
developed by \citet{1999AJ....117.2052C}.  The PCA code outputs 
eigenspectra, eigencoefficients, and repaired data spectra. 

\begin{figure}
\epsscale{0.8}
\plotone{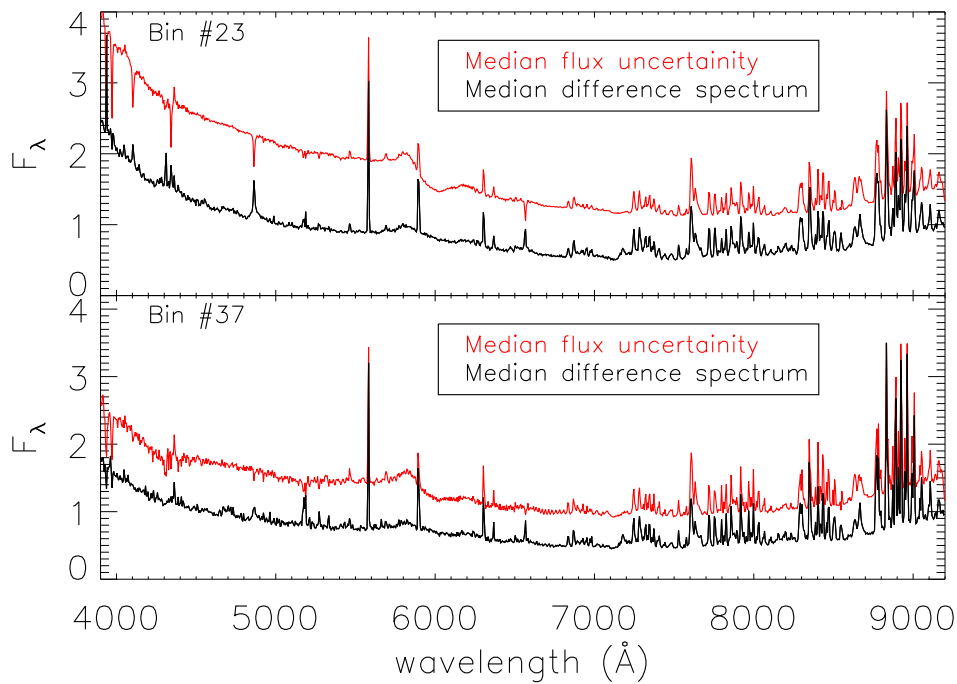}
\caption{
The median difference spectra and median noise spectra for bins \#23 
(top) and \#37 (bottom).  The median difference spectra are black and the 
median noise spectra are red.  The median difference spectra were 
constructed by taking the absolute value of the difference of each 
original spectrum and its reconstructed spectrum, and then taking 
the median of the difference spectra in vacuum wavelength bins of width 
5\AA. We took the median of the noise spectra in similar wavelength 
bins to create the median noise spectra.  We demonstrate in both bins 
that the use of four eigencoefficients reconstructs the original 
spectra to within the measurement noise.}
\label{Fig:MedianDiff}
\end{figure}

\begin{figure}
\epsscale{0.8}
\plotone{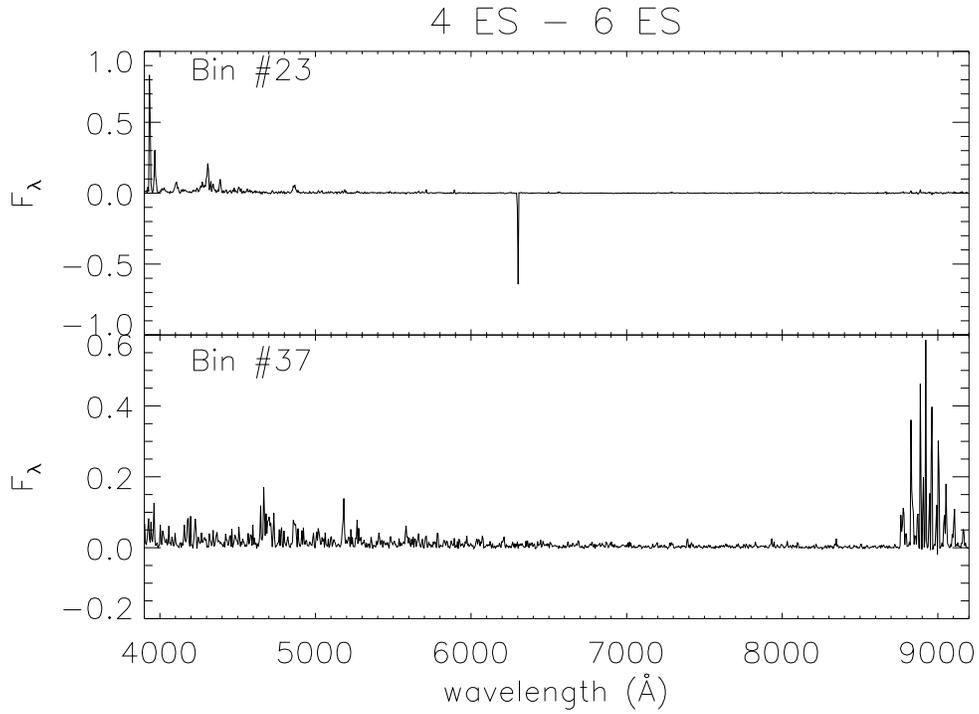}
\caption{A comparison of the median difference spectra for PCA decompositions
based on four and six eigenspectra for bins \#23 (top) and \#37 (bottom).
The median difference spectra were constructed by taking the absolute value 
of the difference of each original spectrum and its reconstructed spectrum,
and then taking the median. The results for PCA decomposition based on four 
eigenspectra are shown in Figure~\ref{Fig:MedianDiff}. These two panels
show the difference between the results from Figure~\ref{Fig:MedianDiff}
and analogous median difference spectra for six eigencomponents.}
\label{Fig:Compare4to6}
\end{figure}

We investigated spectral decompositions using varying numbers of 
eigenspectra and found that the first four eigencomponents are sufficient 
to describe the observed spectra within the measurement noise (for a single
spectrum). 
Figure ~\ref{Fig:MedianDiff} compares the median difference between the original 
and reconstructed spectra with the median noise in the original spectra 
for bins \#23 and \#37.  As shown, the spectra reconstructed using four 
eigencomponents are consistent with the original spectra to well within 
the typical measurement error. Furthermore, Figure ~\ref{Fig:Compare4to6}
demonstrates that the difference between PCA decompositions
based on four and six eigenspectra is minor (a few percent or less over
most of the wavelength range). 

\begin{figure}
\epsscale{0.8}
\plotone{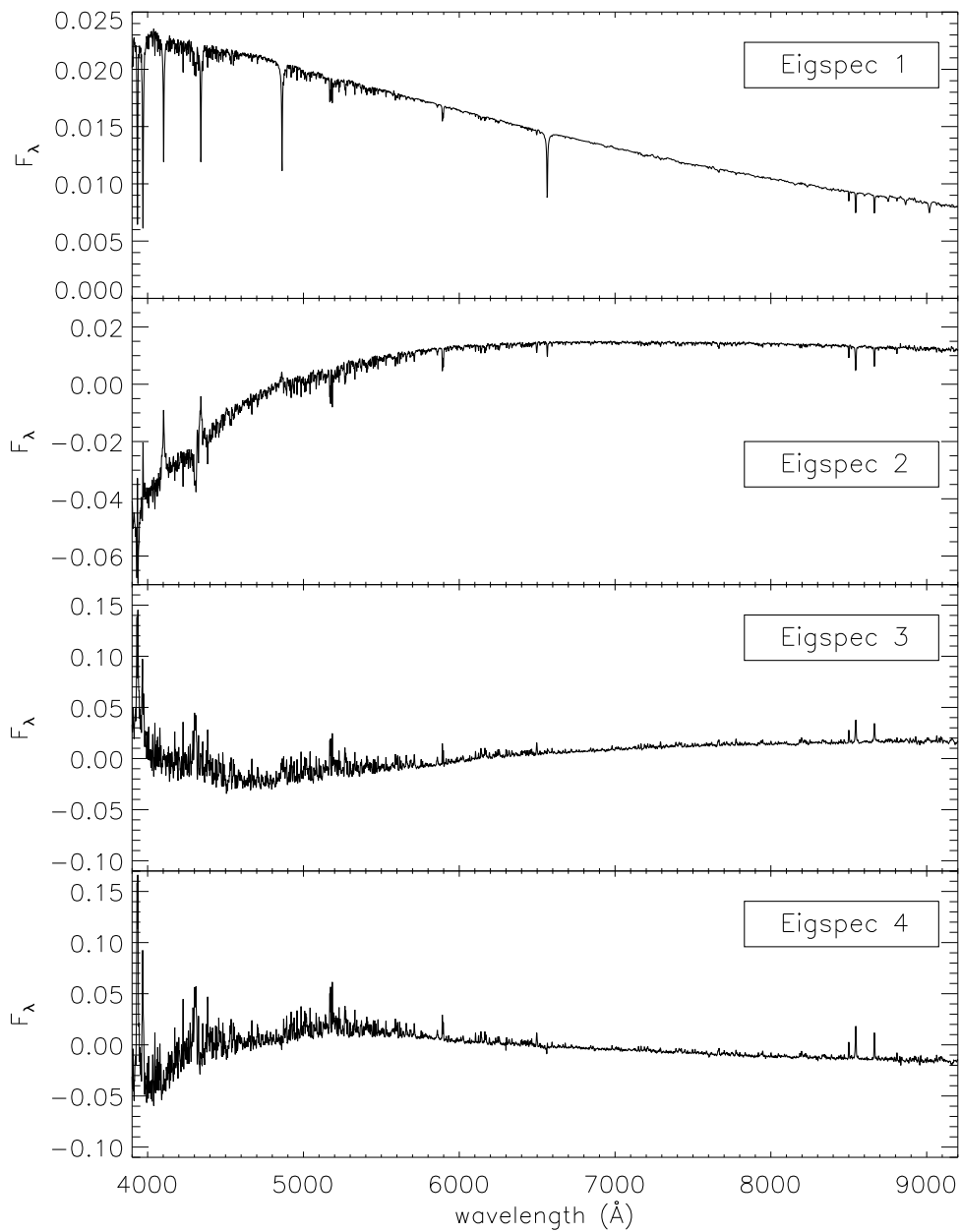}
\epsscale{1.0}
\caption{The four eigenspectra ($F_\lambda$) generated for the 6156 stars 
in bin \#23 ($0.24 < g-r < 0.26$), plotted at vacuum wavelengths. The 
first eigenspectrum closely resembles a metal-poor ($[Fe/H]\sim-1.5$) 
subdwarf. The variation of spectra in this bin is expected to be 
dominated by the variation of stellar metallicity.}
\label{Fig:Bin23ES}
\end{figure}

\begin{figure}
\epsscale{0.8}
\plotone{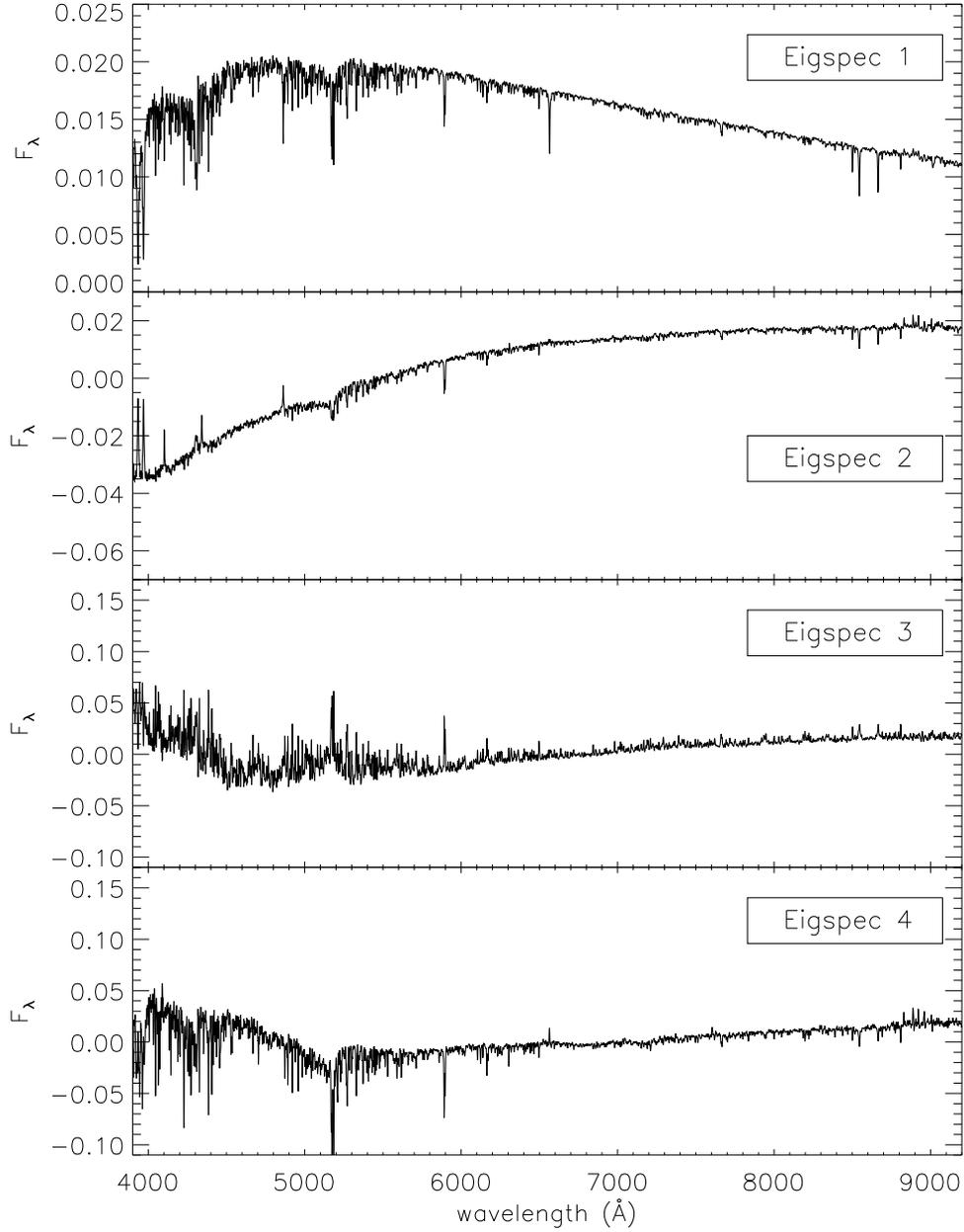}
\epsscale{1.0}
\caption{The four eigenspectra generated for the 8253 stars in bin \#37
($0.52 < g-r < 0.54$), plotted at vacuum wavelenths. The first 
eigenspectrum closely resembles a metal-rich ($[Fe/H]\sim-0.7$) 
dwarf. The variation of spectra in this bin is expected due to 
a high fraction of giant stars (which presumably also have lower 
metallicity than the majority of stars in the sample).}
\label{Fig:Bin37ES}
\end{figure}

This is a much smaller number of eigencomponents than typically required.
For example, \citet{1998MNRAS.298..361B} used the first 10 components
to expand stellar spectra in their sample. The reason for this difference
is that here we individually treat very narrow bins of the $g-r$ color.
In order to describe spectral variations due to metallicity and gravity 
in each bin (i.e., at nearly a constant effective temperature), a large 
number of eigencomponents is not necessary. 
Figures \ref{Fig:Bin23ES} and \ref{Fig:Bin37ES} show the 
eigenspectra for bins \#23 and \#37, respectively (the spectra 
are plotted in vacuum wavelengths).

\section{ Analysis of PC Decomposition Results}

\subsection{ Correlations between eigencoefficients and SSPP parameters } %% 3.1

\begin{figure}
\epsscale{0.8}
\plotone{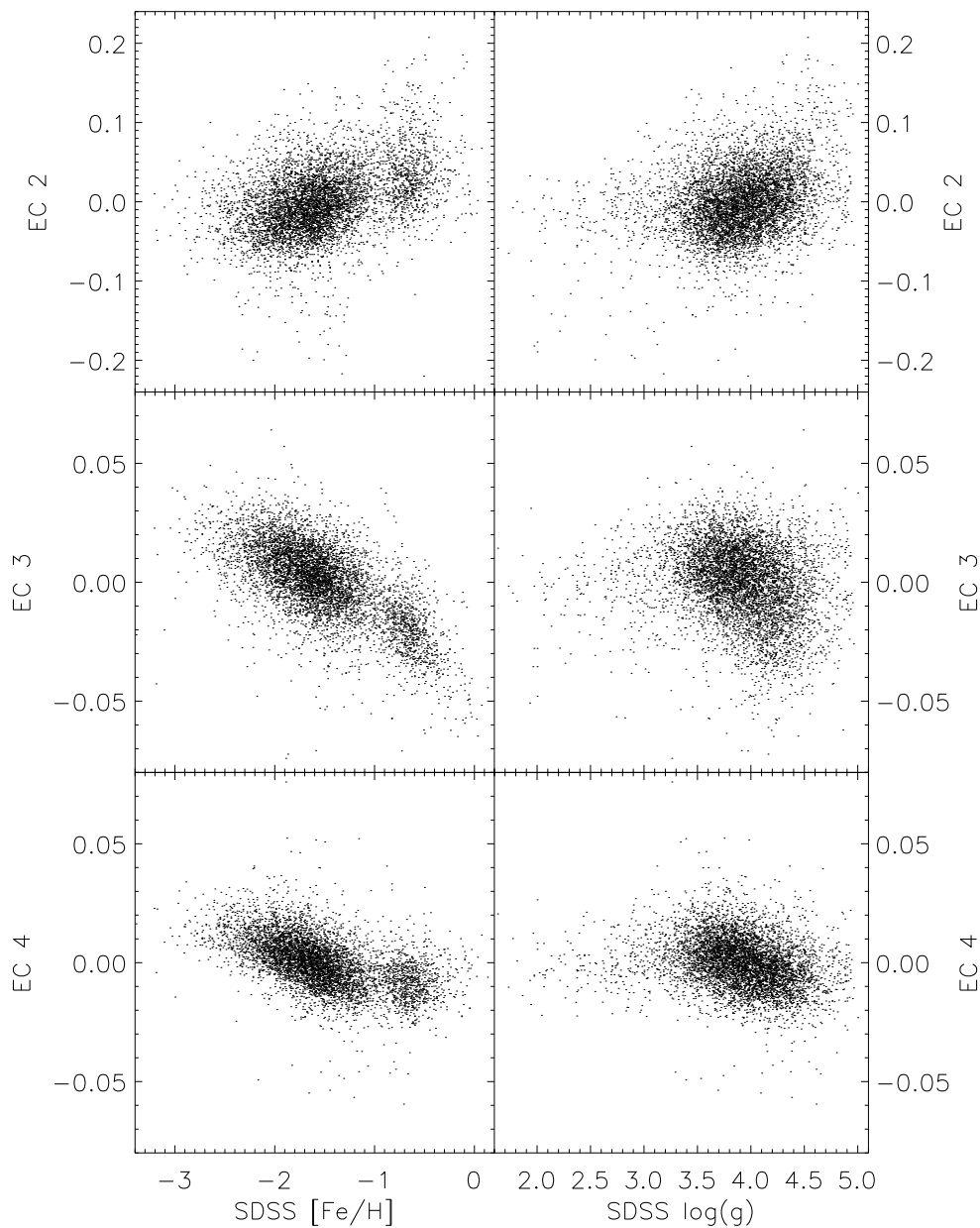}
\epsscale{1.0}
\caption{Eigencoefficients for bin \#23, corresponding to eigenspectra
         shown in Figure~\ref{Fig:Bin23ES}, shown as a function of metallicity 
         (left column) and gravity (right column) computed by SSPP. 
         Note a small fraction of giant stars ($\log(g)<3$) and a correlation
         between metallicity and EC2, EC3, and EC4. The bimodal metallicity 
         distribution reflects the SDSS targeting algorithm and significantly
         different metallicity distributions for halo ($[Fe/H]<-1$) and 
         disk ($[Fe/H]>-1$) stars.}
\label{Fig:Bin23ECpar}
\end{figure}

\begin{figure}
\epsscale{0.8}
\plotone{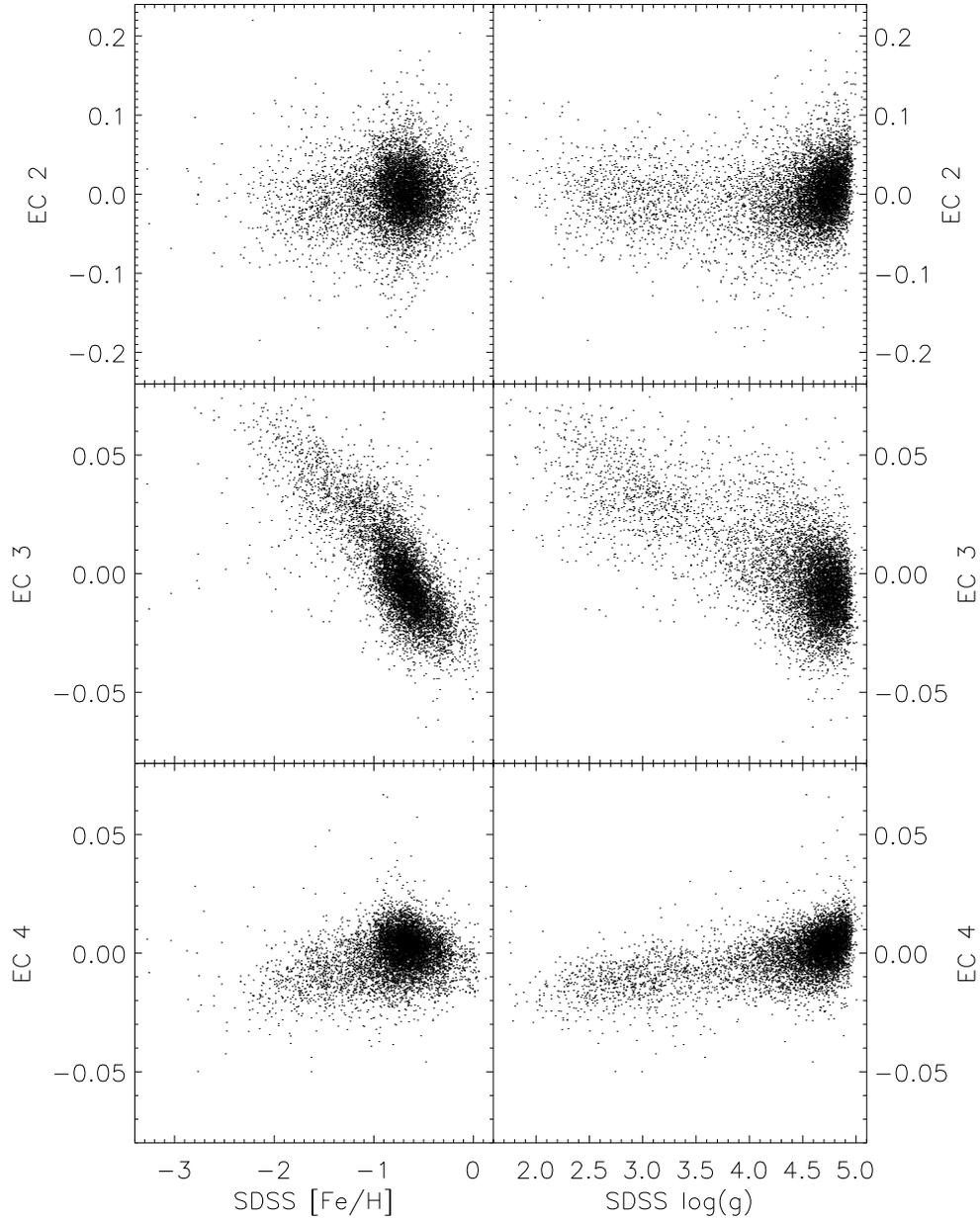}
\epsscale{1.0}
  \caption{Eigencoefficients for bin \#37, corresponding to eigenspectra
         shown in Figure~\ref{Fig:Bin37ES}, shown as a function of metallicity 
         (left column) and gravity (right column) computed by SSPP.
         Note a much larger fraction of giant stars than in Figure~\ref{Fig:Bin23ECpar}.
         The eigencoefficients seem correlated with both metallicity and gravity.
         The sample does not include high-metallicity giants.}
\label{Fig:Bin37ECpar}
\end{figure}

For each $g-r$ bin, the eigencoefficients are expected to encode
information about metallicity and gravity. Figures~\ref{Fig:Bin23ECpar}
and \ref{Fig:Bin37ECpar} demonstrate correlations between PCA 
eigencoefficients (EC) and SSPP metallicity and gravity for
bins \#23 and \#37. For bin \#23, a correlation between metallicity and 
EC2, EC3, and EC4, is present in the data, while there is no correlation
with gravity. Note, however, that this bin includes only a small fraction 
of giant stars (3\% with $\log(g)<3$). For bin \#37, eigencoefficients
are correlated with both metallicity and gravity. However, as demonstrated
in Figure~\ref{Fig:FeHLogg}, the metallicity and gravity are correlated 
for stars in this bin \citep[due to SDSS spectroscopic target selection 
criteria, for more details see][]{Yanny09}, and thus 
it is not clear which parameter is 
driving the correlation with eigencoefficients. A sample of 
high-metallicity giants or low-metallicity dwarfs is required 
to decouple the effects of these two parameters. Such 
stars are not present in the sample; the former are mostly 
nearby disk stars and thus too bright and saturated in SDSS data, while 
the latter are distant halo stars and too faint to be included in the 
SDSS spectroscopic survey \citep{2008arXiv0804.3850I}.

\begin{figure}
\plotone{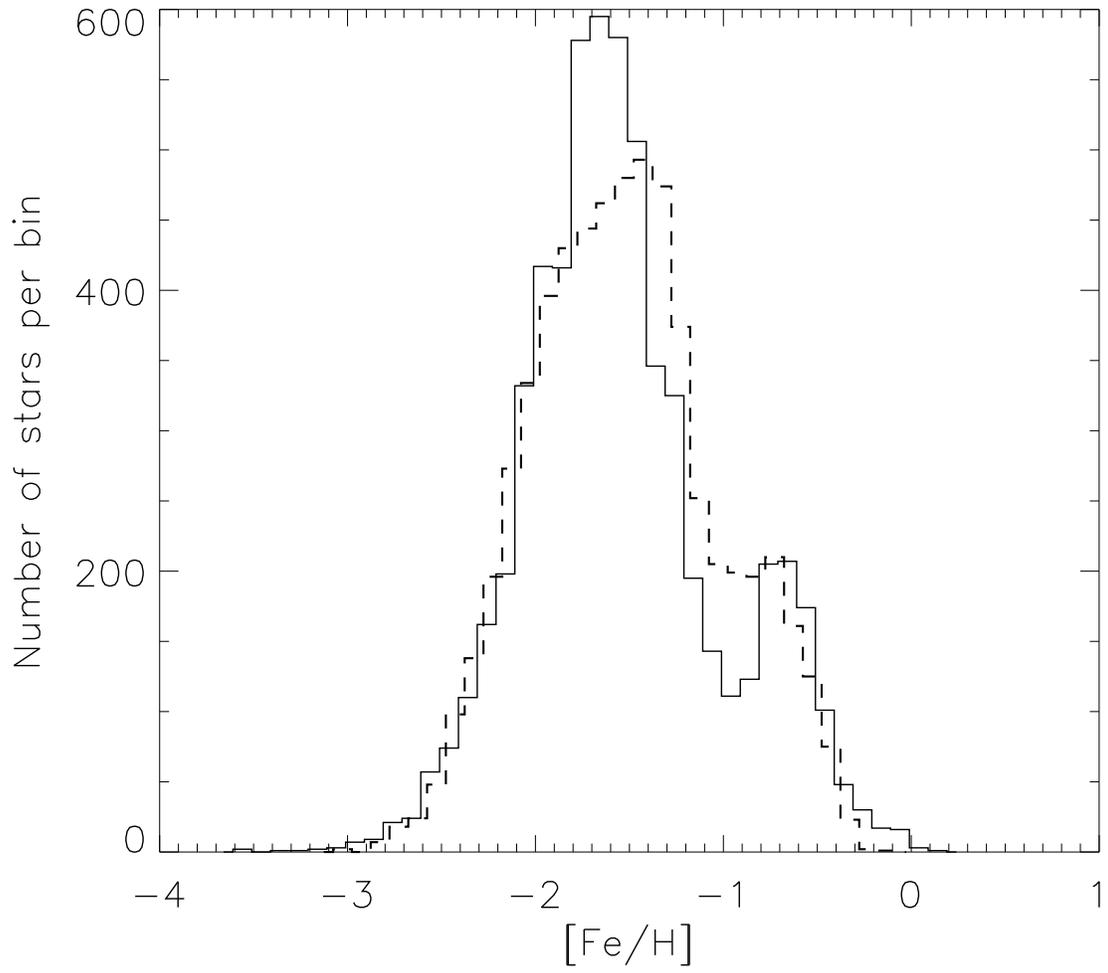}
%\vskip -4.4in
  \caption{A comparison of the metallicity estimates obtained by the 
     SDSS SSPP pipeline (solid histogram) and our PCA-based 
      estimates (dashed histogram).}
  \label{Bin23CompHist}
\end{figure}

For a more quantitative investigation of the correlation observed in 
bin \#23, we fit straight lines to the three 
relationships between metallicity and eigencoefficients 2, 3, 
and 4 (best-fit parabolas lead to the same conclusions). 
We then average the three best-fit metallicity 
estimates obtained using each eigencoefficient and compare the result
to values reported by the SSPP (Figure~\ref{Bin23CompHist}). The resulting
bimodal distributions look similar, though one could argue that 
the values determined with PCA have slightly larger random errors 
(by about 20\%) than the official SDSS values because the distinction 
between the two peaks in metallicity distribution is somewhat erased. 

\begin{figure}
%\vskip -4.4in
\epsscale{0.8}
\plotone{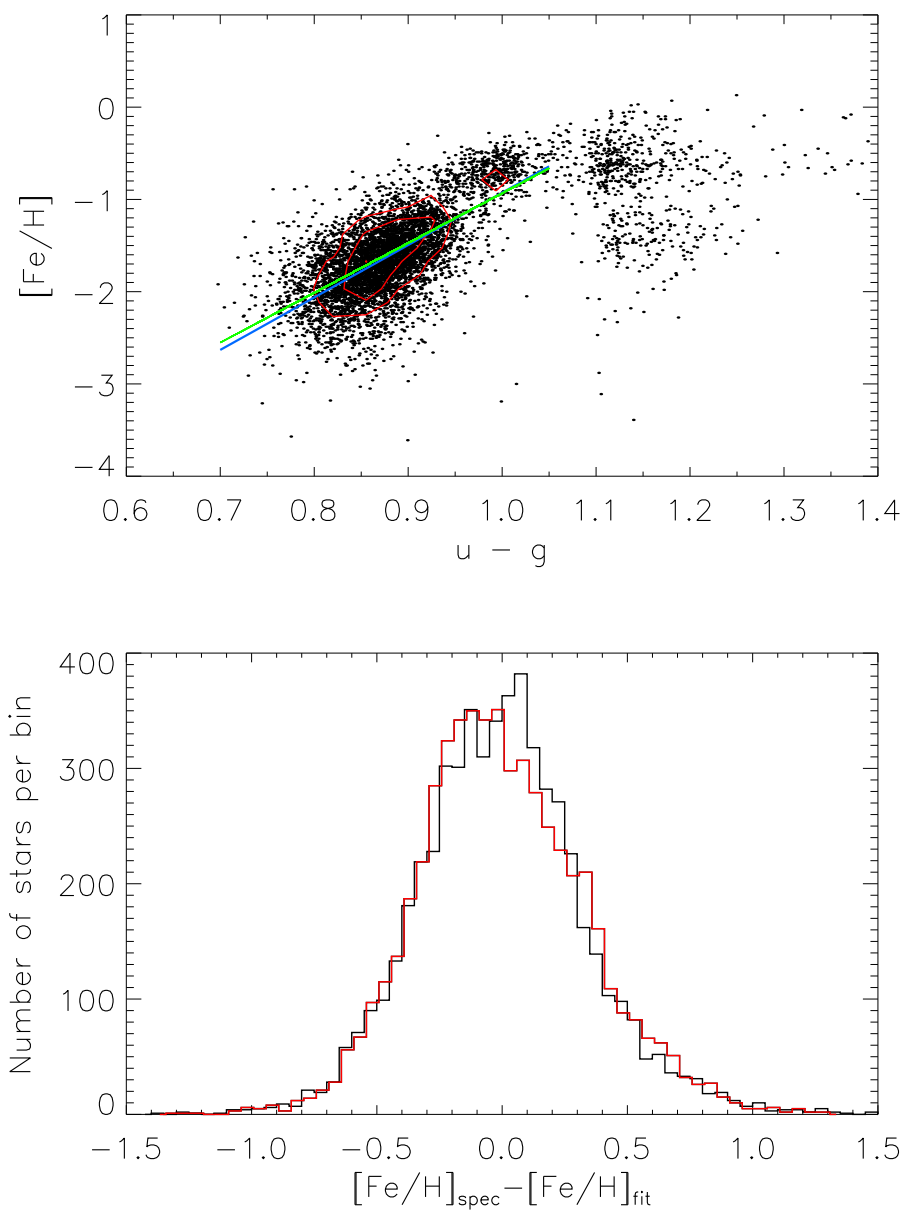}
\caption{An independent test of two metallicity estimators based
on a correlation of metallicity and the $u-g$ color. In the top plot, 
black points illustrate SSPP-measured metallicities, while the red 
contours illustrate our new metallicity measurements.  The green and 
blue lines are the best-fit lines to the two datasets, respectively 
the SSPP fit and the new PCA fit. The bottom plot illustrates the scatter 
of each data set around the best-fit lines, with black being the SSPP 
data and red being the new metallicity data. }
\label{Bin23CompHist2}
\end{figure}

For another comparison of the two metallicity estimators, we use the
$u-g$ color obtained from imaging data. As shown by \citet{2008arXiv0804.3850I}, 
for stars with blue $g-r$ colors, spectroscopic metallicity
is strongly correlated with the $u-g$ color (see the top panel
in Figure~\ref{Bin23CompHist2}). They estimate that the random photometric 
metallicity errors are even smaller than the random metallicity errors 
determined by the SSPP from SDSS spectra (a random $u-g$ error of 0.02 mag 
induces a metallicity error in $[Fe/H]$ that varies from 0.02 dex at 
$[Fe/H]=-0.5$ to 0.11 dex at $[Fe/H]=-1.5$). Hence, for stars from bin \#23, 
we can compare the spectroscopic metallicities determined by PCA and by 
the SSPP using the scatter around an average fit (see the bottom panel 
in Figure~\ref{Bin23CompHist2}; a best-fit parabola leads to the same 
conclusion); the better metallicity determination would have less scatter. 
We find essentially identical error
behavior, which suggests that the parameter precision obtained by PCA 
is comparable to that achieved by the SSPP for stars with $0.2<g-r<0.3$.

\subsection{ Mean Stellar Spectra Determined by PCA} %% 3.2

\begin{figure}
\epsscale{1.02}
\plotone{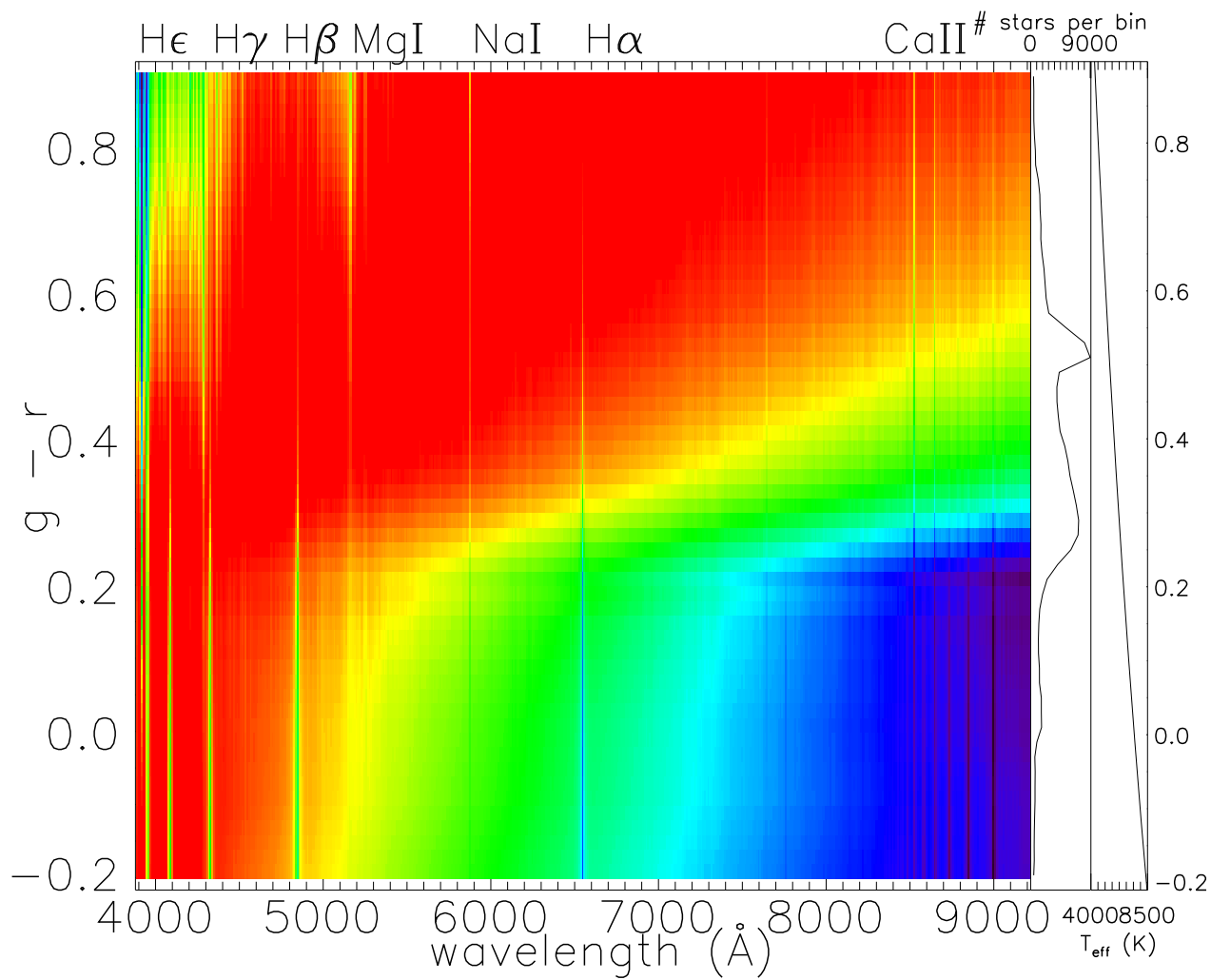}
\epsscale{1.0}
\caption{The progression of mean spectra ($F_\lambda$) plotted in 
vacuum wavelengths as a function of the $g-r$ color determined 
using PCA. The values increase logarithmically from blue to red. 
The two panels on the right side show the mean number of stars 
per 0.02 mag wide $g-r$ bin, and the corresponding median effective 
temperature.  A few spectral absorption lines are marked on top.}
  \label{meanSpec}
\end{figure}

\begin{figure}
\vskip -2.4in
\plotone{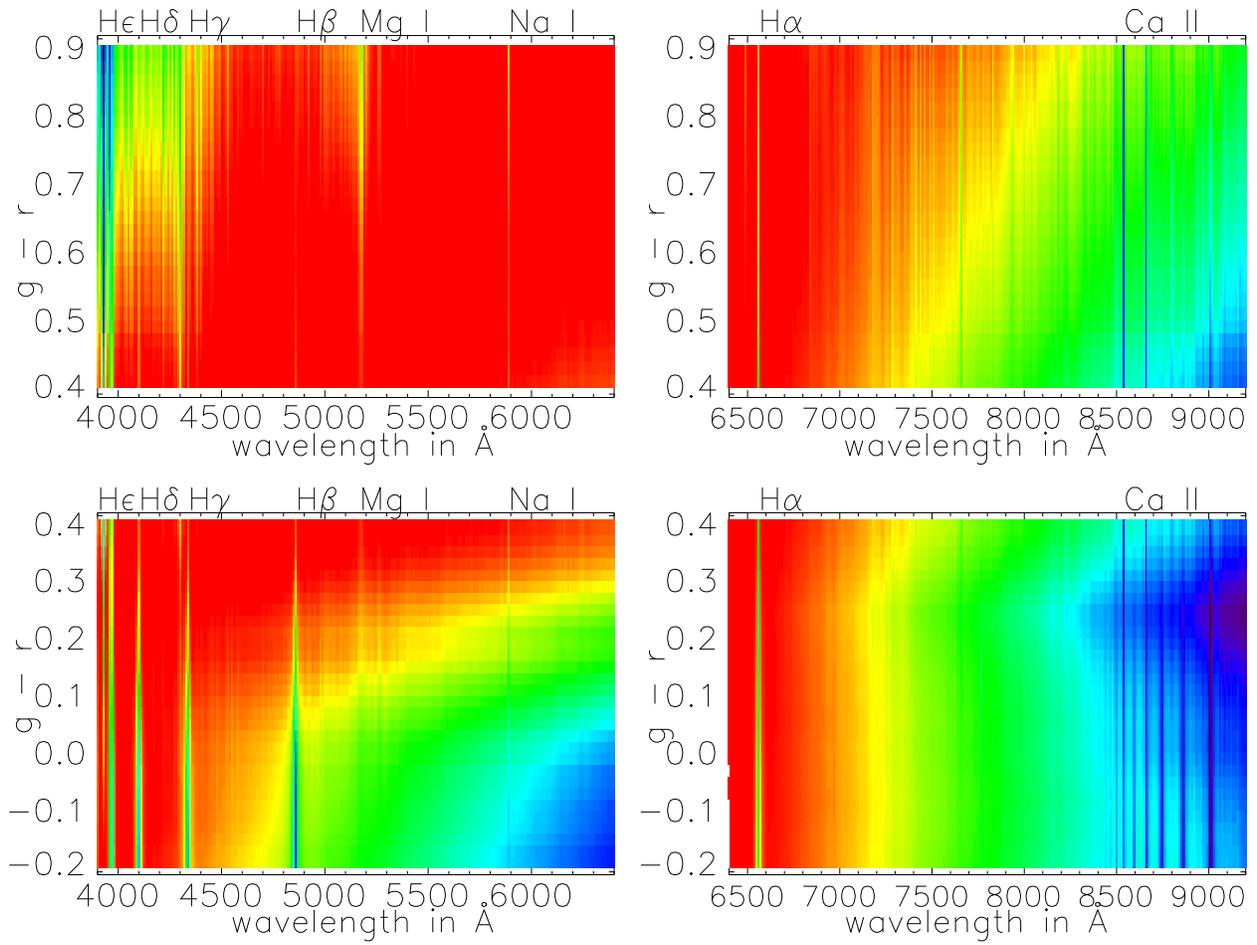}
\caption{Analogous to Figure~\ref{meanSpec}, except that the wavelength
range and the $g-r$ range are split in halves and each panel is separately 
color-coded to increase the contrast of spectral features.} 
\label{comp4}
\end{figure}

One of the PCA products derived in this work is a set of high signal-to-noise 
gap-repaired mean spectra at each $g-r$ color. We show a stack of 55 such spectra in 
Figures~\ref{meanSpec} and \ref{comp4}. The variation of absorption
line strengths and overall continuum shape with the $g-r$ color
are easily discernible. For example, the depth of the $H\alpha$ line
steadily decreases as the $g-r$ color becomes redder. 

In addition to mean spectra that roughly correspond 
to stars with {\it median} metallicity and gravity in a given $g-r$ bin,  
the remaining eigenspectra can be used to generate high signal-to-noise spectra 
for {\it any} combination of basic stellar parameters (effective temperature,
metallicity, and gravity), within the parameter space probed by SDSS. 
Hence, given an arbitrary spectrum, one can attempt a low-dimensional
fit using our library of eigenspectra, which we make publicly 
available\footnote{http://www.astro.washington.edu/users/ivezic/rmcgurk/PCApublic.shtml}.

\subsection{ Variation of Mean Stellar Spectra with Metallicity and Gravity }%% 3.3

\begin{figure}
\plotone{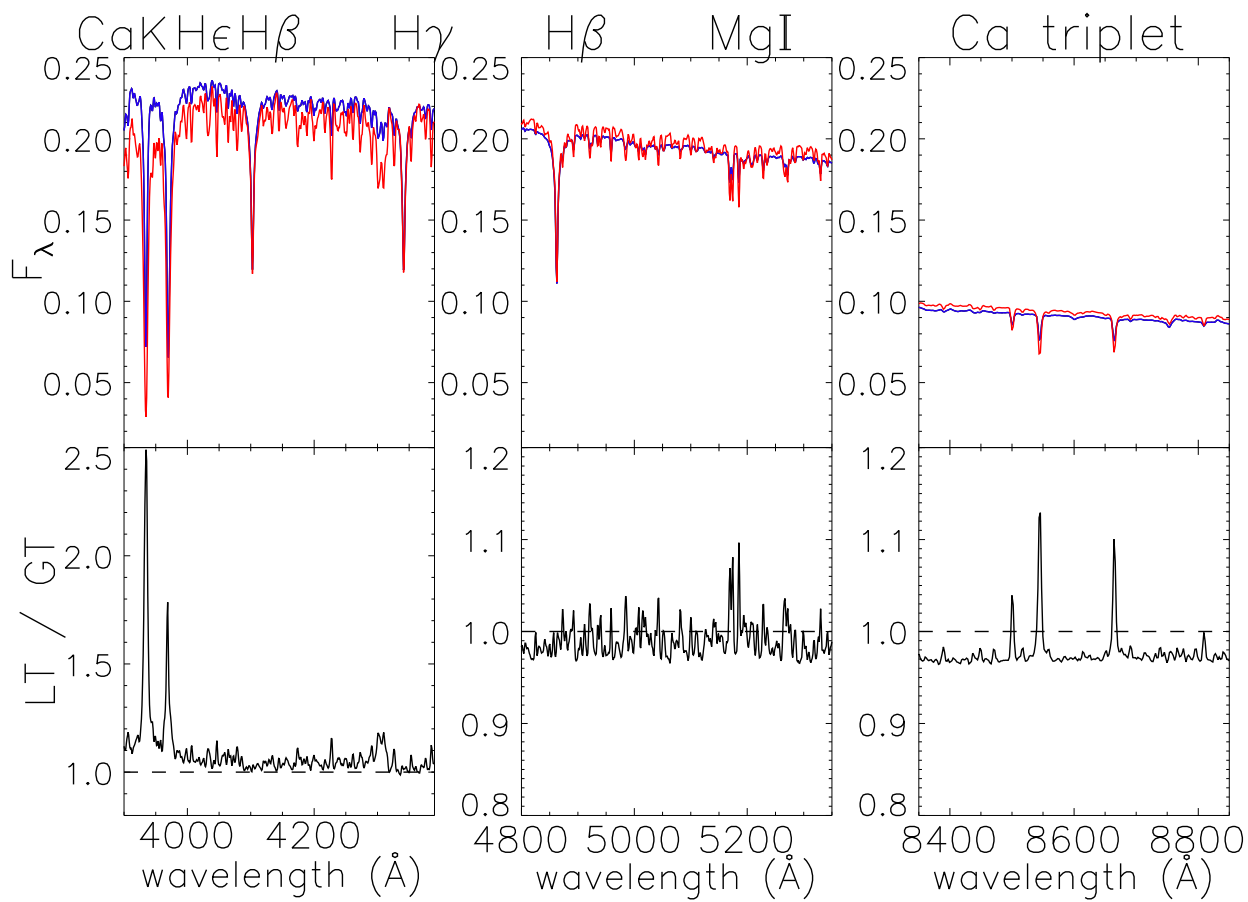}
  \caption{The top panels show two reconstructed spectra in three characteristic
vacuum wavelength ranges for bin \#23. The spectra are generated using mean 
  eigencoefficients of the two clumps separated by $[Fe/H]=-1.05$ (see 
Fig.~\ref{Fig:Bin23ECpar}; the blue line corresponds to the clump with $[Fe/H] < -1.05$
and the red line to the more metal-rich clump). The bottom panels show
the ratio of the low-metallicity ($[Fe/H] \sim -1.5$) and the high-metallicity 
spectrum ($[Fe/H] \sim -0.6$).}
\label{Fig:Bin23Recon}
\end{figure}

\begin{figure}
\plotone{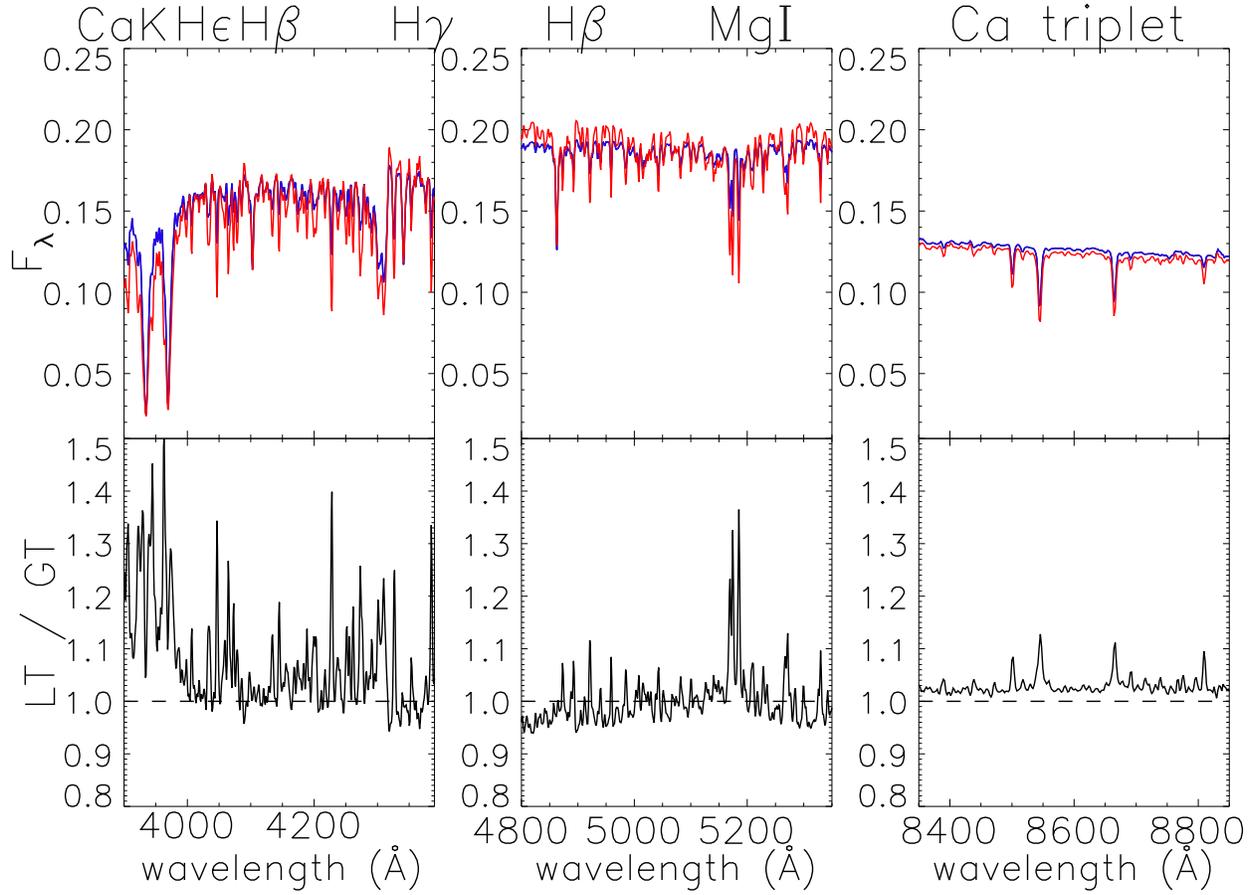}
  \caption{Similar to Figure~\ref{Fig:Bin23Recon}, except that spectra 
   are reconstructed for bin \#37. The two reconstructed spectra are 
   generated using mean eigencoefficients for the clump with $log(g)<3.7$ 
    (blue line), and the clump with $log(g)>3.7$ (red line), see 
     Fig.~\ref{Fig:Bin37ECpar} for the distribution of eigencoefficients.}
\label{Fig:Bin37Recon}
\end{figure}

The mean spectra shown in Figures~\ref{meanSpec} and \ref{comp4}
are averaged over metallicity and gravity distributions in the
corresponding color bins. The small variations of stellar spectra 
with metallicity and gravity, at a fixed $g-r$ color, can be 
studied by drawing eigencoefficients from their observed distribution
in a given bin. Given the quality of SDSS spectra and the high
signal-to-noise ratio of the mean spectra (due to large number
of stars per bin), these variations can be studied in great detail. 
An example of such a study is shown in Figures~\ref{Fig:Bin23Recon}
and \ref{Fig:Bin37Recon}, where we contrast low- and high-metallicity
stars from bin \#23, and low (giants) and high (dwarfs) log(g) stars
from bin \#37. For example, the sensitivity of the Ca triplet
(around $\sim$8600 \AA) to both metallicity and gravity is easily
discernible (this wavelength range is exploited by the RAVE and Gaia
surveys).

\section{      Summary       }

We have applied Principal Component Analysis (PCA) to $\sim$100,000 
stellar spectra obtained by the SDSS.  After binning the sample using
the $g-r$ color to study line variation at a nearly constant effective
temperature as a function of the metallicity and gravity, we find that 
the first four eigenspectra fully capture
the observed spectral variations in each bin (within the noise in 
individual SDSS spectra). We analyze correlations 
between our PCA eigencoefficients and SEGUE Stellar Parameters Pipeline (SSPP)
metallicity, and then use these correlations to measure metallicity.  
We find similar performance between the PCA-measured metallicity and 
the SSPP metallicity.  This similarity suggests that {\it random errors 
of SSPP parameters are just about as small as the signal-to-noise ratios 
of SDSS spectra allow.}

We make publicly available the resulting high signal-to-noise mean 
spectra and the other three eigenspectra for all 55 color bins. These data 
can be used to generate high quality spectra for an arbitrary combination of 
effective temperature, metallicity, and gravity, within the parameter space 
probed by SDSS. The utility of such spectra is wide and varied. Most obvious
applications include searching for unusual spectra, training of various 
spectral classification methods, and color synthesis in arbitrary optical 
photometric systems, as well as various educational programs.

%%%%%%%%%%%%%%%%%%%%%%%%%%%%%%%%%%%%%%%%%%%%%%%%%%%%%%%%%%%%%%%%%%%%%%%%%%%%%%%%%
\vskip 0.4in \leftline{Acknowledgments}

We are thankful to Andy Connolly and Ching-Wa Yip for making their PCA code
available to us, and to Robert Lupton for suggesting this project. We 
acknowledge support by NSF grants AST-615991 and AST-0707901. 

Funding for the SDSS and SDSS-II has been provided by the Alfred P. Sloan
Foundation, the Participating Institutions, the National Science Foundation, 
the U.S. Department of Energy, the National Aeronautics and Space 
Administration, the Japanese Monbukagakusho, the Max Planck Society, and 
the Higher Education Funding Council for England. The SDSS Web Site is 
http://www.sdss.org/.

The SDSS is managed by the Astrophysical Research Consortium for the 
Participating Institutions. The Participating Institutions are the 
American Museum of Natural History, Astrophysical Institute Potsdam, 
University of Basel, University of Cambridge, Case Western Reserve 
University, University of Chicago, Drexel University, Fermilab, the 
Institute for Advanced Study, the Japan Participation Group, Johns 
Hopkins University, the Joint Institute for Nuclear Astrophysics, the 
Kavli Institute for Particle Astrophysics and Cosmology, the Korean 
Scientist Group, the Chinese Academy of Sciences (LAMOST), Los Alamos 
National Laboratory, the Max-Planck-Institute for Astronomy (MPIA), 
the Max-Planck-Institute for Astrophysics (MPA), New Mexico State 
University, Ohio State University, University of Pittsburgh, University 
of Portsmouth, Princeton University, the United States Naval Observatory, 
and the University of Washington.

\bibliographystyle{apj}
\bibliography{refs}

\begin{thebibliography}{28}
\expandafter\ifx\csname natexlab\endcsname\relax\def\natexlab#1{#1}\fi

\bibitem[{{Abazajian} \& {Sloan Digital Sky
  Survey}(2008)}]{Abazajian.2008arXiv}
{Abazajian}, K., \& {Sloan Digital Sky Survey}, f.~t. 2008, ArXiv e-prints

\bibitem[{{Allende Prieto} {et~al.}(2006){Allende Prieto}, {Beers}, {Wilhelm},
  {Newberg}, {Rockosi}, {Yanny}, \& {Lee}}]{2006ApJ...636..804A}
{Allende Prieto}, C., {Beers}, T.~C., {Wilhelm}, R., {Newberg}, H.~J.,
  {Rockosi}, C.~M., {Yanny}, B., \& {Lee}, Y.~S. 2006, \apj, 636, 804

\bibitem[{{Allende Prieto} {et~al.}(2007){Allende Prieto}, {Sivarani}, {Beers},
  {Lee}, {Koesterke}, {Shetrone}, {Sneden}, {Lambert}, {Wilhelm}, {Rockosi},
  {Lai}, {Yanny}, {Ivans}, {Johnson}, {Aoki}, {Bailer-Jones}, \& {Re
  Fiorentin}}]{2007arXiv0710.5780A}
{Allende Prieto}, C., {et~al.} 2007, ArXiv e-prints, 710

\bibitem[{{Bailer-Jones} {et~al.}(1998){Bailer-Jones}, {Irwin}, \& {von
  Hippel}}]{1998MNRAS.298..361B}
{Bailer-Jones}, C.~A.~L., {Irwin}, M., \& {von Hippel}, T. 1998, \mnras, 298,
  361

\bibitem[{{Beers} {et~al.}(2006){Beers}, {Lee}, {Sivarani}, {Allende Prieto},
  {Wilhelm}, {Fiorentin}, {Bailer-Jones}, {Norris}, \& {the SEGUE Calibration
  Team}}]{2006MmSAI..77.1171B}
{Beers}, T.~C., {et~al.} 2006, Memorie della Societa Astronomica Italiana, 77,
  1171

\bibitem[{{Connolly} \& {Szalay}(1998)}]{1998IAUS..179..376C}
{Connolly}, A.~J., \& {Szalay}, A.~S. 1998, in IAU Symposium, Vol. 179, New
  Horizons from Multi-Wavelength Sky Surveys, ed. B.~J. {McLean}, D.~A.
  {Golombek}, J.~J.~E. {Hayes}, \& H.~E. {Payne}, 376--+

\bibitem[{{Connolly} \& {Szalay}(1999)}]{1999AJ....117.2052C}
{Connolly}, A.~J., \& {Szalay}, A.~S. 1999, \aj, 117, 2052

\bibitem[{{Connolly} {et~al.}(1995){Connolly}, {Szalay}, {Bershady}, {Kinney},
  \& {Calzetti}}]{1995AJ....110.1071C}
{Connolly}, A.~J., {Szalay}, A.~S., {Bershady}, M.~A., {Kinney}, A.~L., \&
  {Calzetti}, D. 1995, \aj, 110, 1071

\bibitem[{{Covey} {et~al.}(2007){Covey}, {Ivezi{\'c}}, {Schlegel},
  {Finkbeiner}, {Padmanabhan}, {Lupton}, {Ag{\"u}eros}, {Bochanski}, {Hawley},
  {West}, {Seth}, {Kimball}, {Gogarten}, {Claire}, {Haggard}, {Kaib},
  {Schneider}, \& {Sesar}}]{Covey.2007AJ}
{Covey}, K.~R., {et~al.} 2007, \aj, 134, 2398

\bibitem[{{Eisenstein} {et~al.}(2001){Eisenstein}, {Annis}, {Gunn}, {Szalay},
  {Connolly}, {Nichol}, {Bahcall}, {Bernardi}, {Burles}, {Castander},
  {Fukugita}, {Hogg}, {Ivezi{\'c}}, {Knapp}, {Lupton}, {Narayanan}, {Postman},
  {Reichart}, {Richmond}, {Schneider}, {Schlegel}, {Strauss}, {SubbaRao},
  {Tucker}, {Vanden Berk}, {Vogeley}, {Weinberg}, \&
  {Yanny}}]{Eisenstein.2001AJ}
{Eisenstein}, D.~J., {et~al.} 2001, \aj, 122, 2267

\bibitem[{{Eisenstein} {et~al.}(2006){Eisenstein}, {Liebert}, {Harris},
  {Kleinman}, {Nitta}, {Silvestri}, {Anderson}, {Barentine}, {Brewington},
  {Brinkmann}, {Harvanek}, {Krzesi{\'n}ski}, {Neilsen}, {Long}, {Schneider}, \&
  {Snedden}}]{2006ApJS..167...40E}
---. 2006, \apjs, 167, 40

\bibitem[{{Fan}(1999)}]{1999AJ....117.2528F}
{Fan}, X. 1999, \aj, 117, 2528

\bibitem[{{Finlator} {et~al.}(2000){Finlator}, {Ivezi{\'c}}, {Fan}, {Strauss},
  {Knapp}, {Lupton}, {Gunn}, {Rockosi}, {Anderson}, {Csabai}, {Hennessy},
  {Hindsley}, {McKay}, {Nichol}, {Schneider}, {Smith}, {York}, \& {the SDSS
  Collaboration}}]{2000AJ....120.2615F}
{Finlator}, K., {et~al.} 2000, \aj, 120, 2615

\bibitem[{{Gunn} {et~al.}(2006){Gunn}, {Siegmund}, {Mannery}, {Owen}, {Hull},
  {Leger}, {Carey}, {Knapp}, {York}, {Boroski}, {Kent}, {Lupton}, {Rockosi},
  {Evans}, {Waddell}, {Anderson}, {Annis}, {Barentine}, {Bartoszek}, {Bastian},
  {Bracker}, {Brewington}, {Briegel}, {Brinkmann}, {Brown}, {Carr},
  {Czarapata}, {Drennan}, {Dombeck}, {Federwitz}, {Gillespie}, {Gonzales},
  {Hansen}, {Harvanek}, {Hayes}, {Jordan}, {Kinney}, {Klaene}, {Kleinman},
  {Kron}, {Kresinski}, {Lee}, {Limmongkol}, {Lindenmeyer}, {Long}, {Loomis},
  {McGehee}, {Mantsch}, {Neilsen}, {Neswold}, {Newman}, {Nitta}, {Peoples},
  {Pier}, {Prieto}, {Prosapio}, {Rivetta}, {Schneider}, {Snedden}, \&
  {Wang}}]{2006AJ....131.2332G}
{Gunn}, J.~E., {et~al.} 2006, \aj, 131, 2332

\bibitem[{{Ivezi{\'c}} {et~al.}(2008{\natexlab{a}}){Ivezi{\'c}}, {Sesar},
  {Juric}, {Bond}, {Dalcanton}, {Rockosi}, {Yanny}, {Newberg}, {Beers},
  {Allende Prieto}, {Wilhelm}, {Lee}, {Sivarani}, {Norris}, {Bailer-Jones}, {Re
  Fiorentin}, {Schlegel}, {Uomoto}, {Lupton}, {Knapp}, {Gunn}, {Covey}, {Allyn
  Smith}, {Miknaitis}, {Doi}, {Tanaka}, {Fukugita}, {Kent}, {Finkbeiner},
  {Munn}, {Pier}, {Quinn}, {Hawley}, {Anderson}, {Kiuchi}, {Chen}, {Bushong},
  {Sohi}, {Haggard}, {Kimball}, {Barentine}, {Brewington}, {Harvanek},
  {Kleinman}, {Krzesinski}, {Long}, {Nitta}, {Snedden}, {Lee}, {Harris},
  {Brinkmann}, {Schneider}, \& {York}}]{2008arXiv0804.3850I}
{Ivezi{\'c}}, {\v Z}., {et~al.} 2008{\natexlab{a}}, ArXiv e-prints, 804

\bibitem[{{Ivezi{\'c}} {et~al.}(2008{\natexlab{b}}){Ivezi{\'c}}, {Tyson},
  {Allsman}, {Andrew}, {Angel}, \& {for the LSST
  Collaboration}}]{2008arXiv0805.2366I}
{Ivezi{\'c}}, {\v Z}., {Tyson}, J.~A., {Allsman}, R., {Andrew}, J., {Angel},
  R., \& {for the LSST Collaboration}. 2008{\natexlab{b}}, ArXiv e-prints, 805

\bibitem[{{Ivezi{\'c}} {et~al.}(2005){Ivezi{\'c}}, {Vivas}, {Lupton}, \&
  {Zinn}}]{2005AJ....129.1096I}
{Ivezi{\'c}}, {\v Z}., {Vivas}, A.~K., {Lupton}, R.~H., \& {Zinn}, R. 2005,
  \aj, 129, 1096

\bibitem[{{Lee} {et~al.}(2007{\natexlab{a}}){Lee}, {Beers}, {Sivarani},
  {Allende Prieto}, {Koesterke}, {Wilhelm}, {Norris}, {Bailer-Jones}, {Re
  Fiorentin}, {Rockosi}, {Yanny}, {Newberg}, \& {Covey}}]{2007arXiv0710.5645L}
{Lee}, Y.~S., {et~al.} 2007{\natexlab{a}}, ArXiv e-prints, 710

\bibitem[{{Lee} {et~al.}(2007{\natexlab{b}}){Lee}, {Beers}, {Sivarani},
  {Johnson}, {An}, {Wilhelm}, {Allende Prieto}, {Koesterke}, {Re Fiorentin},
  {Bailer-Jones}, {Norris}, {Yanny}, {Rockosi}, {Newberg}, {Cudworth}, \&
  {Pan}}]{2007arXiv0710.5778L}
---. 2007{\natexlab{b}}, ArXiv e-prints, 710

\bibitem[{{Lenz} {et~al.}(1998){Lenz}, {Newberg}, {Rosner}, {Richards}, \&
  {Stoughton}}]{1998ApJS..119..121L}
{Lenz}, D.~D., {Newberg}, J., {Rosner}, R., {Richards}, G.~T., \& {Stoughton},
  C. 1998, \apjs, 119, 121

\bibitem[{{Richards} {et~al.}(2002){Richards}, {Fan}, {Newberg}, {Strauss},
  {Vanden Berk}, {Schneider}, {Yanny}, {Boucher}, {Burles}, {Frieman}, {Gunn},
  {Hall}, {Ivezi{\'c}}, {Kent}, {Loveday}, {Lupton}, {Rockosi}, {Schlegel},
  {Stoughton}, {SubbaRao}, \& {York}}]{Richards.2002AJ}
{Richards}, G.~T., {et~al.} 2002, \aj, 123, 2945

\bibitem[{{Schlegel} {et~al.}(1998){Schlegel}, {Finkbeiner}, \&
  {Davis}}]{1998ApJ...500..525S}
{Schlegel}, D.~J., {Finkbeiner}, D.~P., \& {Davis}, M. 1998, \apj, 500, 525

\bibitem[{{Sirko} {et~al.}(2004){Sirko}, {Goodman}, {Knapp}, {Brinkmann},
  {Ivezi{\'c}}, {Knerr}, {Schlegel}, {Schneider}, \&
  {York}}]{2004AJ....127..899S}
{Sirko}, E., {et~al.} 2004, \aj, 127, 899

\bibitem[{{Smol{\v c}i{\'c}} {et~al.}(2004){Smol{\v c}i{\'c}}, {Ivezi{\'c}},
  {Knapp}, {Lupton}, {Pavlovski}, {Iliji{\'c}}, {Schlegel}, {Smith}, {McGehee},
  {Silvestri}, {Hawley}, {Rockosi}, {Gunn}, {Strauss}, {Fan}, {Eisenstein}, \&
  {Harris}}]{2004ApJ...615L.141S}
{Smol{\v c}i{\'c}}, V., {et~al.} 2004, \apjl, 615, L141

\bibitem[{{Strauss} {et~al.}(2002){Strauss}, {Weinberg}, {Lupton}, {Narayanan},
  {Annis}, {Bernardi}, {Blanton}, {Burles}, {Connolly}, {Dalcanton}, {Doi},
  {Eisenstein}, {Frieman}, {Fukugita}, {Gunn}, {Ivezi{\'c}}, {Kent}, {Kim},
  {Knapp}, {Kron}, {Munn}, {Newberg}, {Nichol}, {Okamura}, {Quinn}, {Richmond},
  {Schlegel}, {Shimasaku}, {SubbaRao}, {Szalay}, {Vanden Berk}, {Vogeley},
  {Yanny}, {Yasuda}, {York}, \& {Zehavi}}]{Strauss.2002AJ}
{Strauss}, M.~A., {et~al.} 2002, \aj, 124, 1810

\bibitem[{{Yanny} {et~al.}(2009){Yanny}, {Rockosi}, {Newberg}, {Knapp},
  {Adelman-McCarthy}, {Alcorn}, {Allam}, {Allende Prieto}, {An}, {Anderson},
  {Anderson}, {Bailer-Jones}, {Bastian}, {Beers}, {Bell}, {Belokurov},
  {Bizyaev}, {Blythe}, {Bochanski}, {Boroski}, {Brinchmann}, {Brinkmann},
  {Brewington}, {Carey}, {Cudworth}, {Evans}, {Evans}, {Gates}, {G{\"a}nsicke},
  {Gillespie}, {Gilmore}, {Gomez-Moran}, {Grebel}, {Greenwell}, {Gunn},
  {Jordan}, {Jordan}, {Harding}, {Harris}, {Hendry}, {Holder}, {Ivans},
  {Ivezi{\v c}}, {Jester}, {Johnson}, {Kent}, {Kleinman}, {Kniazev},
  {Krzesinski}, {Kron}, {Kuropatkin}, {Lebedeva}, {Lee}, {Leger}, {L{\'e}pine},
  {Levine}, {Lin}, {Long}, {Loomis}, {Lupton}, {Malanushenko}, {Malanushenko},
  {Margon}, {Martinez-Delgado}, {McGehee}, {Monet}, {Morrison}, {Munn},
  {Neilsen}, {Nitta}, {Norris}, {Oravetz}, {Owen}, {Padmanabhan}, {Pan},
  {Peterson}, {Pier}, {Platson}, {Fiorentin}, {Richards}, {Rix}, {Schlegel},
  {Schneider}, {Schreiber}, {Schwope}, {Sibley}, {Simmons}, {Snedden}, {Smith},
  {Stark}, {Stauffer}, {Steinmetz}, {Stoughton}, {Subba Rao}, {Szalay},
  {Szkody}, {Thakar}, {Thirupathi}, {Tucker}, {Uomoto}, {Vanden Berk},
  {Vidrih}, {Wadadekar}, {Watters}, {Wilhelm}, {Wyse}, {Yarger}, \&
  {Zucker}}]{Yanny09}
{Yanny}, B., {et~al.} 2009, \aj, 137, 4377

\bibitem[{{Yip} {et~al.}(2004){Yip}, {Connolly}, {Vanden Berk}, {Ma},
  {Frieman}, {SubbaRao}, {Szalay}, {Richards}, {Hall}, {Schneider}, {Hopkins},
  {Trump}, \& {Brinkmann}}]{2004AJ....128.2603Y}
{Yip}, C.~W., {et~al.} 2004, \aj, 128, 2603

\bibitem[{{York} {et~al.}(2000){York}, {Adelman}, {Anderson}, {Anderson},
  {Annis}, {Bahcall}, {Bakken}, {Barkhouser}, {Bastian}, {Berman}, {Boroski},
  {Bracker}, {Briegel}, {Briggs}, {Brinkmann}, {Brunner}, {Burles}, {Carey},
  {Carr}, {Castander}, {Chen}, {Colestock}, {Connolly}, {Crocker}, {Csabai},
  {Czarapata}, {Davis}, {Doi}, {Dombeck}, {Eisenstein}, {Ellman}, {Elms},
  {Evans}, {Fan}, {Federwitz}, {Fiscelli}, {Friedman}, {Frieman}, {Fukugita},
  {Gillespie}, {Gunn}, {Gurbani}, {de Haas}, {Haldeman}, {Harris}, {Hayes},
  {Heckman}, {Hennessy}, {Hindsley}, {Holm}, {Holmgren}, {Huang}, {Hull},
  {Husby}, {Ichikawa}, {Ichikawa}, {Ivezi{\'c}}, {Kent}, {Kim}, {Kinney},
  {Klaene}, {Kleinman}, {Kleinman}, {Knapp}, {Korienek}, {Kron}, {Kunszt},
  {Lamb}, {Lee}, {Leger}, {Limmongkol}, {Lindenmeyer}, {Long}, {Loomis},
  {Loveday}, {Lucinio}, {Lupton}, {MacKinnon}, {Mannery}, {Mantsch}, {Margon},
  {McGehee}, {McKay}, {Meiksin}, {Merelli}, {Monet}, {Munn}, {Narayanan},
  {Nash}, {Neilsen}, {Neswold}, {Newberg}, {Nichol}, {Nicinski}, {Nonino},
  {Okada}, {Okamura}, {Ostriker}, {Owen}, {Pauls}, {Peoples}, {Peterson},
  {Petravick}, {Pier}, {Pope}, {Pordes}, {Prosapio}, {Rechenmacher}, {Quinn},
  {Richards}, {Richmond}, {Rivetta}, {Rockosi}, {Ruthmansdorfer}, {Sandford},
  {Schlegel}, {Schneider}, {Sekiguchi}, {Sergey}, {Shimasaku}, {Siegmund},
  {Smee}, {Smith}, {Snedden}, {Stone}, {Stoughton}, {Strauss}, {Stubbs},
  {SubbaRao}, {Szalay}, {Szapudi}, {Szokoly}, {Thakar}, {Tremonti}, {Tucker},
  {Uomoto}, {Vanden Berk}, {Vogeley}, {Waddell}, {Wang}, {Watanabe},
  {Weinberg}, {Yanny}, \& {Yasuda}}]{York.etal.2000AJ}
{York}, D.~G., {et~al.} 2000, \aj, 120, 1579

\end{thebibliography}

\newpage

\end{document}